# Mapping fast evolution of transient surface photovoltage dynamics using G-Mode Kelvin probe force microscopy

*Liam Collins,†,‡ Mahshid Ahmadi,ζ Jiajun Qin,ζ Olga S. Ovchinnikova,†,‡ Bin Hu,ζ Stephen Jesse†,‡ and Sergei V. Kalinin,†,‡*

† Center for Nanophase Materials Sciences, Oak Ridge National Laboratory, Oak Ridge, Tennessee 37831, USA

‡ Institute for Functional Imaging of Materials, Oak Ridge National Laboratory, Oak Ridge, Tennessee 37831, USA

ζ Joint Institute for Advanced Materials, Department of Materials Science and Engineering, University of Tennessee, Knoxville, Tennessee 37996, USA

**ABSTRACT.** Optoelectronic phenomena in materials such as organic/inorganic hybrid perovskites depend on a complex interplay between light induced carrier generation and fast (electronic) and slower (ionic) processes, all of which are known to be strongly affected by structural inhomogeneities such as interfaces and grain boundaries. Here, we develop a time resolved Kelvin probe force microscopy (KPFM) approach, based on the G-Mode SPM platform, allowing quantification of surface photovoltage (SPV) with microsecond temporal and nanoscale spatial resolution. We demonstrate the approach on methylammonium lead bromide (MAPbBr3) thin films and further highlight the usefulness of unsupervised clustering methods to quickly discern spatial variability in the information rich SPV dataset. Using this technique, we observe concurrent spatial and ultra-fast temporal variations in the SPV generated across the thin film, indicating that structure is likely responsible for the heterogenous behavior.

**INTRODUCTION**

Solar energy is critical towards meeting the ever increasing global energy demands, necessitating the continuous search for novel photovoltaic materials and device structures. Organic/inorganic hybrid perovskites (OIHPs) solar cells have gained significant attention with an impressive power conversion efficiency of 22.1%.[1] Beyond this, methylammonium lead bromide (MAPbBr$_3$) has attracted wide interest as an active material in light emitting diodes (LEDs)[2] and as an ideal candidate in tandem structures with a Si or a Cu(InGa)Se$_2$ solar cell.[3] This is due to its superior charge transport with high and nearly balanced carrier mobility for both electrons and holes, large light absorption coefficient, strong photoluminescent quantum efficiency, large and tunable band gap, low defect density as well as solution processability and low cost.[4-6]

Despite tremendous advancement in OIHP optoelectronic device performances in the last decade, a full understanding of several observed physical behaviors, including coupled fast and slow relaxation time scales,[7-9] hysteretic transport behavior,[10] and non-uniform optoelectronic characteristics[11-14] remain largely elusive. Many material challenges remain unsolved, and it is widely accepted that optimization and improved longevity of OIHP devices requires precise knowledge of the charge distribution behavior at local inhomogeneity's (e.g. grain to grain variability, grain boundaries, interfaces and, and defects). [15-16]

While scanning probe microscopies (SPM) are ideally suited for spatial probing of materials structure and functional properties on these length scales, they are inherently slow, restricting applications to equilibrium or very slow processes (e.g. > seconds). Of particular relevance for photovoltaics, Kelvin probe force microscopy (KPFM)[17], is a non-invasive mode of AFM which allows the simultaneous mapping of the topography and the local contact potential difference (CPD) between tip and sample with nanometer resolution. The CPD value measured by KPFM is related to work function for metal samples, surface potential in dielectrics, band bending in semiconductors and surface photovoltage (SPV) in photovoltaics.[18-19] The ability to correlate nanoscale structural and electronic/electrochemical characteristics has made KPFM a powerful method to study electronic devices including solar cells and LEDs.[20-24] In particular, spatially resolved SPV measurements can be related to important characteristics including carrier diffusion length[20-21] carrier lifetime and local recombination rates,[25] characteristics which are fundamental to understanding charge generation process in photovoltaic materials.[26] At the same time, capturing information on these processes requires KPFM techniques which are capable of probing both fast (ns-µs) and slow (ms-s) processes.

The factors limiting the measurement bandwidth in classical KPFM are the lock-in amplifier time constant and the bandwidth of the bias feedback loop, well below the mechanical bandwidth of the AFM cantilever itself. Practically, measuring SPV using KPFM involves capturing the CPD or surface potential ($V_{sp}$) under both illuminated ($V_{sp\text{-light}}$) and dark ($V_{sp\text{-dark}}$) conditions, where SPV=$V_{sp\text{-light}}$ - $V_{sp\text{-dark}}$. However, a KPFM image under either condition can take several 10s of minutes to capture, and hence, the accuracy of the SPV measurement depends on the stability of the tip-surface potential in the time frame of the entire measurement. Furthermore, the long measurement time makes studies of SPV under different environmental or illumination conditions (wavelength, intensity, and light soaking/degradation effect) impractical.[27] This is especially important in the study of materials including OIHPs which involve ion migration.[28]

As the result, in KPFM measurements all information on dynamic processes below the measurement timescale (e.g. ms) is lost. At the same time, a complete understanding of the role of trapped charges and ion migration play in anomalous observations such as the hysteretic behavior and light soaking effect in OIHPs would benefit greatly from time-resolved (tr) SPV measurements using KPFM. Indeed, the KPFM community have tried to respond in recent years by developing tr-Electrostatic force microscopy[30-34] and tr-KPFM[27,35] approaches for investigation of time dependent optoelectronic properties. However, these approaches have some limitations including either; lacking quantitative information on local potentials,[30-33] sometimes operated in single point mode (i.e. no spatial contrast),[27, 33] or where quantitative KPFM imaging has been realized the time resolution is still limited to ~16 seconds.[35]

Recently we developed the G-Mode acquisition approach for SPM measurements,[36] and subsequently developed open loop G-Mode KPFM[37-38] as a method of probing surface potentials on microsecond timescales, well below the mechanical bandwidth of the cantilever.[39] Here we combine G-Mode KPFM with photoexcitation as a method to map SPV dynamics of photovoltaic samples with temporal properties. As a proof of principle, we choose a thin films of methylammonium lead bromide (MAPbBr$_3$) on ITO/PEDOT:PSS substrate. We further highlight the usefulness of adopting unsupervised clustering algorithms for quickly and effectively discerning local deviations in optoelectronic properties from the high dimensional and information rich datasets afforded by G-Mode KPFM.

**RESULTS AND DISCUSSION**

Figure 1 describes the measurement setup. All KPFM measurements were operated in lift mode, or dual pass mode, in which the sample topography is recorded in the first pass, and the KPFM measurement is performed during a second pass at a predefined distance above the sample (100 nm unless otherwise stated). The G-Mode platform was used to capture the photodetector signal at high sampling rates (~ 2-4 MHz) as the tip was raster scanned (scan rate ~ 0.4 Hz) over the sample in lift mode. For G-Mode KPFM, a sinusoidal voltage is applied to the conductive cantilever generating a dynamic electrostatic force between tip and sample. The dynamic cantilever response due to the electrostatic force is encoded in the photodetector signal. De-noising of the photodetector signal in G-Mode KFPM is realized in a post processing step using Fourier filtering (Noise thresholding and low pass filter) on the entire data, parsed into individual line segments. We note that the knowledge of the noise floor and the capability to inspect the data allows for adaptable post-experiment filtering, not traditionally afforded in laboratory settings. Further, G-Mode KPFM allows flexibility in exploration of frequency filter and noise threshold settings without effecting the original raw dataset. [40]

Once the data has been processed, recovery of the true electrostatic force is achieved through deconvolution of the cantilever transfer function (calibrated at the beginning of the measurement) using the Fast Free Force (F³R) reconstruction method outlined previously.[39] In the case of EFM or KPFM, the tip-sample electrostatic force ($F_{es}$), established between the grounded sample and conductive probe is written as:

$$F_{es} = \frac{1}{2}C'(V_{tip} - V_{SP})^2 \qquad \text{Eq. (1)}$$

Where $C'$ is the tip-sample capacitance gradient, $V_{sp}$ is the surface potential or more precisely the CPD between tip and sample. The tip voltage is $V_{tip}=V_{dc}+V_{ac}\cos(\omega t)$; however, application of a $V_{dc}$ bias offset is not a requirement in G-Mode KPFM and in principle any arbitrary waveform can be adopted. As seen from Eq (1), the recovered electrostatic force is expected to have parabolic voltage dependence. The quantitative values of $V_{sp}$ can be determined by fitting the functional form of the force vs voltage relation. Correspondingly, the readout rate of the $V_{sp}$ is governed by the time per period of oscillation of the tip voltage. After functional fitting of the recovered force, the data matrix comprises a multidimensional dataset of $V_{sp}$ ($x$, $y$, time). Although not considered in this work, the functional form of the force vs bias relationship can be further related to information on the capacitance (i.e. second order fit parameter), as well as charging events or polarization effects (i.e. deviation for purely parabolic response) which can be inferred from the functional form of the response.

For SPV measurements the excitation laser is periodically modulated on and off to induce a photo response in the sample. The laser excitation waveform can be configured in multiple ways; here we chose to include a single illumination event per pixel (4.096 ms) such that the light was modulated on (2.048 ms) and off (~2.048 ms) at each spatial location. The SPV was calculated by subtracting the $V_{sp}$ recorded under illuminated conditions from the $V_{sp}$ under dark conditions.

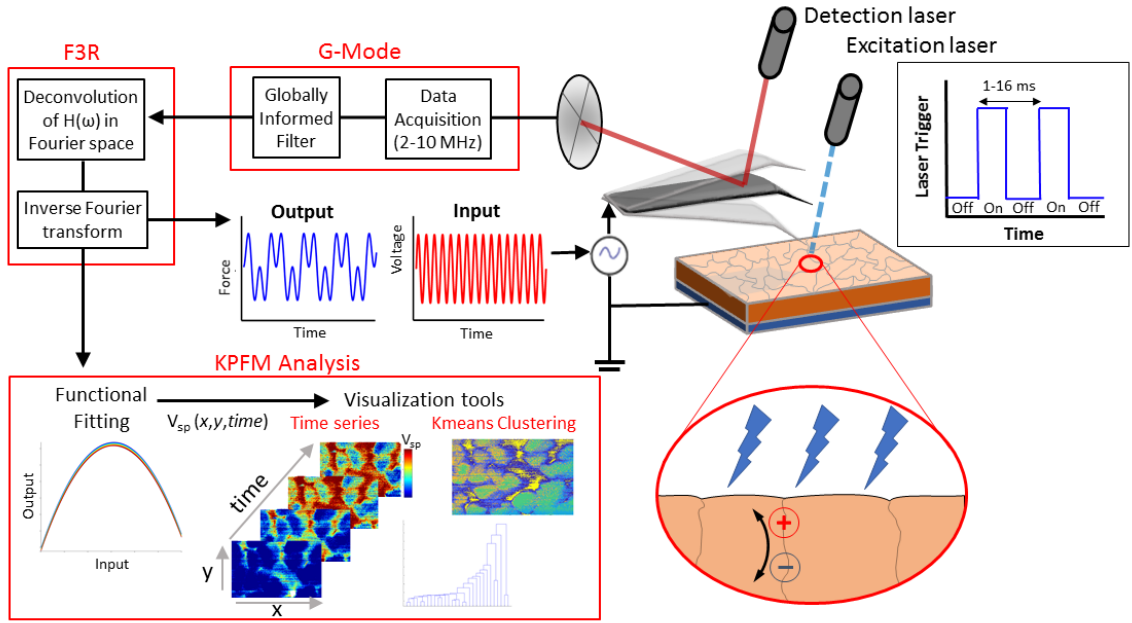

Figure 1. Schematic of the measurement set-up for surface photovoltage measurements by G-Mode KPFM.

Shown in Figure 2(a) is the topography height profile of a representative area on the MAPbBr$_3$ thin film sample, having grains of ~1-2 µm in size. The KPFM surface potential maps show variations in $V_{sp}$ at grain boundaries, within grain facets, and in defective regions, as shown in Figure 2(b). The heterogeneous variation in surface potential at grain boundaries or within single facets could be a result of any number of effects including local doping, or chemical segregation upon crystallization of the film, presence of defect states or shallow trapping levels or even creation of small polaronic states due to localized lattice strain and molecular orientations.[13-14, 41-42] However, the focus of this work is not to determine the precise origin of such variation, instead we attempt to capture of characteristic light induced photovoltage behavior by developing a time resolved approach.

KPFM measurements under dark and illuminated conditions were performed on a smaller area of the sample as shown in Figure 2(c). Using standard KPFM, it was found that upon illumination (see Figure 2(f)), a decrease in the measured $V_{sp}$ was observed relative to the $V_{sp}$ measured under dark conditions (see Figure 2(e)), indicating a reduction of the work function of the MAPbBr$_3$. The calculated SPV is shown in Figure 2(d), demonstrating an average SPV value of -78 ± 12 mV. Local grain to grain variations in the SPV can be observed, as well as variations in SPV in areas correlating with grain boundaries. In addition, the negative SPV (Figure 2(d)) indicates a downward band bending due to accumulation of negative charges at the surface. Meanwhile, a gradual positive drift in SPV can be seen in the direction of the scan (bottom to top), this can be indicative of a slow (>>sec) relaxation process within the material likely a result of excess charge relaxation through the film thickness, ion migration or the relaxation of trapped charges.[35] Notably, while KPFM captures the time averaged processes and is suited for probing such slow processes, these measurements provide little to no information on fast processes taking place below the KPFM measurement time (~11 min).

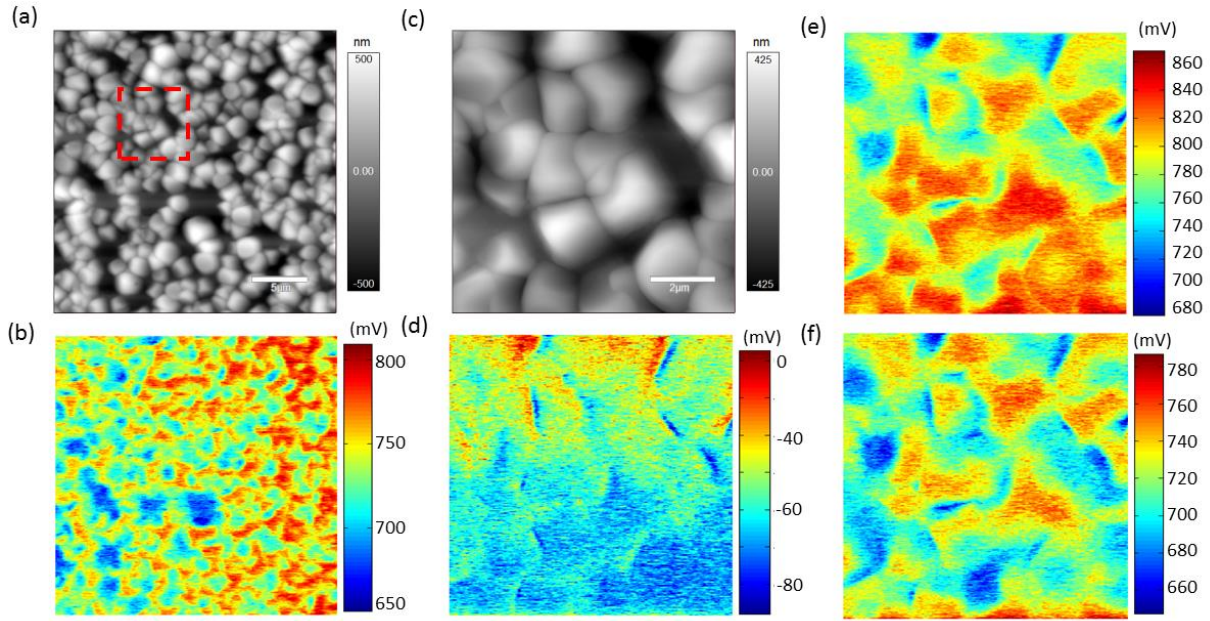

Figure 2. KPFM surface photovoltage measurement on MAPbBr$_3$ thin film. (a,b) Topography and surface potential of a 25 μm$^2$ region. (c,d) Topography and SPV in a smaller region indicated by a red box in (a). SPV was calculated from subtraction of the $V_{sp}$ measured under (f) dark and (e) illuminated conditions.

To explore the fast (μs – ms) light-induced SPV dynamics in MAPbBr$_3$ thin film, we utilize G-Mode KPFM. As a first representation, the time averaged SPV data is shown (Figure 3(b)), determined by subtracting the mean SPV value recorded during the illuminated (Figure 3(c)) and dark (Figure 3(d)) states for each pixel. In agreement with Figure 2, we see a decrease in the surface potential upon illumination. However, for G-Mode KPFM we observe a much larger SPV values than that observed in standard KPFM (-140 ± 28 mV vs -78 ± 18 mV). This is likely due to differences in measurement timescales between G-Mode KPFM and classical KPFM that allow the observations of light-induced fast processes prior the onset of ionic screening, see Figure 2(d). Furthermore, the SPV contrast at the grain boundaries was found to be larger and more pronounced in G-Mode KPFM than in standard KPFM. This result could also be related to the fact that in G-Mode KPFM measurements, the illumination pulse width is much shorter than for KPFM (2 ms vs 18 min) which may be below the time it takes for appreciable charge relaxation to be observed. Indeed, it has been suggested that prolonged illumination led to trap filling by photogenerated carriers, pronounced ion migration/redistribution and structural changes.[43-45] In addition, the light soaking can enhance built in potential which will influence the measured SPV. [43]

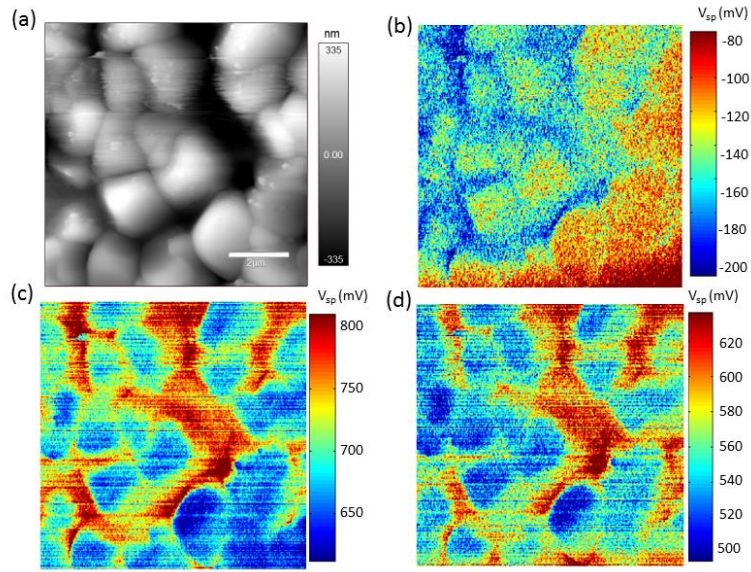

Figure 3. G-Mode KPFM surface photovoltage measurement on MAPbBr$_3$ thin film. (a) Topography height and (b) SPV determined from the time averaged (~2 ms) $V_{sp}$ recorded under (c) dark and (d) illuminated conditions.

In Figure 4 a time series depicting the evolution of the SPV immediately after turning off (see Figure 4(a)) and on illumination (see Figure 4(b)) is shown. When the laser is turned on and off, the SPV decays and rises with a time scale that depends on the physics of the processes controlling the excess charge density. From evolution of decay after light off the contribution of free and trapped charges can be identified as free carriers usually decay faster and trapped carrier and ion migration happen at much slower rates. Therefore, analysis of fast surface photovoltage evolution by G-Mode KPFM can give rise information on photogenerated charge transport. In the G-Mode KPFM measurements shown here, the drive voltage was chosen to coincide with the resonance frequency (70.5 KHz) to achieve resonance enhancement. The drive frequency, or more precisely the period of oscillation, in turn determines the time resolution of the CPD readout. In this case, it was ~14.9 μs. On closer inspection of the time series, subtle differences in spatio-temporal SPV measurements can be observed, as shown in Figure 4. Previously in organic bulk heterojunction films, the SPV decay showed both slow (~minutes to hours) and fast (~ms or faster) component which was attributed to the trapping and detrapping of deep level defects and recombination of excess free carriers, respectively.[46] As can be seen in Figure 4(b), we noticed the evolution of fast decay in MAPbBr$_3$ thin films, with most of the photovoltage dissipating on time scales of ~μs. Indeed, photoinduced surface charges can quickly recombine at the surface because of the fast surface recombination rate due to increased density of mobile charges in the timescale detectable by G-Mode KPFM, whereas bulk recombination which happens on a much longer timescale is accessible by KPFM. In this way, combination of both methods may allow separation of different charge relaxation/recombination rates with nanoscale spatial resolution. It should be noted that the evolution in SPV can be also strongly influenced by the surface chemistry of MAPbBr$_3$ thin film as well as the effect of environmental gases and water layers. However, identification of the precise origin of the photovoltage and relaxation is beyond the scope of this manuscript. At the same time, the capability of fast imaging by G-Mode KPFM can ultimately benefit probing the origin of non-homogeneity and transient dynamics of optoelectronic phenomena in this class of complicated semiconductors.

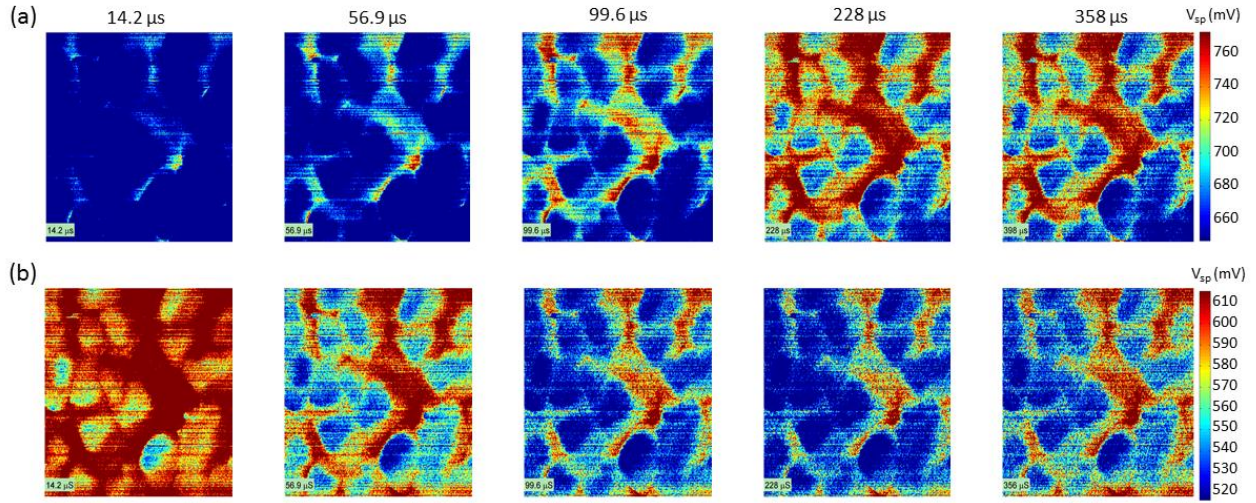

Figure 4. Time series of the G-Mode KPFM surface potential immediately after turning off (a) and (b) on the illumination.

Owing to the high dimensionality and information rich data recoded using G-Mode KPFM, more sophisticated visualization tools will be valuable for meaningful interpretation of the data. Here, we demonstrate the benefit of such approaches by adopting K-means clustering to more clearly visualize the spatio-temporal response of the SPV. First, to determine the appropriate number of clusters to be considered we use the elbow method.[47] Generally speaking, this method looks at the percentage of variance explained as a function of the number of clusters, as seen in Figure 5(a). Clearly, the difference in variance explained by cluster 3 and 4 is quite large, however, as the number of clusters increases >5, the marginal gain in variance explained decreases leading to a curvature of the plot. For this dataset, the optimal number of clusters was determined to be 6. The spatial grouping of the clusters is shown in Figure 5(b). This cluster map can be correlated with the sample structure, including grain boundaries ($k=1$), variance within and grain facets ($k=2-5$), and defective areas ($k=6$). Importantly, this method allows comparison of the mean response within each cluster, as shown in Figure 5(c). From inspection of the cluster distribution and characteristic SPV dynamics, the defects (cluster 6) are shown to result in a reduced SPV (~12 mV). Surprisingly, the SPV response for grain boundaries and certain grain facets captured in cluster 3 and 4 were also shown to demonstrate a reduction in the SPV value (~15 mV) compared to grains in clusters 1, 2 and 5. This complexity in functionality validates the use of G-Mode KPFM as well as highlighting the importance of adopting data science analysis tools.

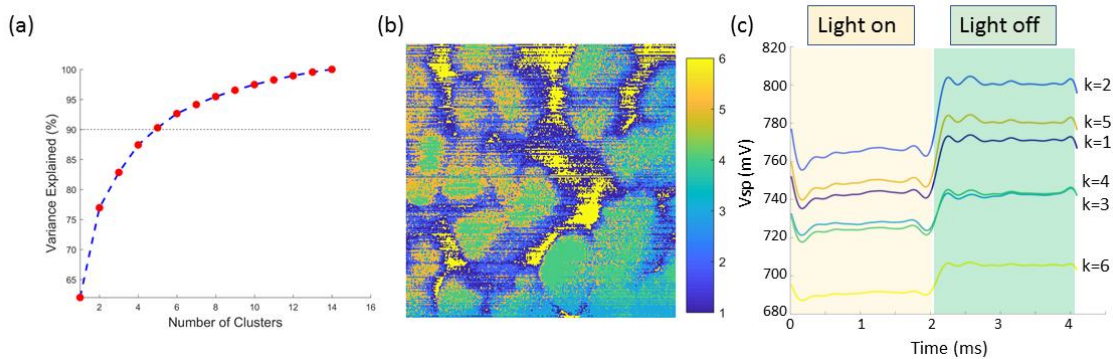

Figure 5. Unsupervised clustering of the 3D (*x,y,time*) $V_{sp}$ data provided by G-Mode KPFM. (a) Plot of variance (expressed as a percentage) versus number of clusters. Elbow method was used to determine the lowest number of clusters (*k*=6) which can be used to explain >90 % of the total variance. (b) Clusters and (c) mean response vs cluster location determined using K-means method.

**CONCLUSION**

In this work we have developed a nanoscale imaging approach for quantitative SPV measurements. We have demonstrated its usefulness for probing fast surface photovoltage dynamics in complex optoelectronic material system like OIHPs with mixed ionic and electronic conductivity. We have demonstrated noticeable differences in the surface photovoltage measured using the traditional KPFM approach and G-Mode KPFM approach which can access fast material response dynamics as well as avoiding relaxation processes associated with prolonged exposure. Indeed, SPV signal from KPFM is just the time average of the near equilibrated SPV measured under either illuminated and in dark condition. The time between these two states can be resolved by G-Mode KPFM which can be used to reveal the mechanism of any fast processes in the system. The information rich data afforded by this method will be welcomed across the fields of organic and inorganic solar cell research, as well as the broader application to novel optoelectronic materials, for quantifying the temporal and spatial variance in optoelectronic properties. Importantly, implementation of this technique does not require expensive peripheral devices, or sophisticated triggering circuitry,[30] necessitating only a data acquisition card synced with an arbitrary waveform generator, as such is readily implementable on any AFM platform. We further note, a characteristic feature of this approach is the retention of the original raw data set, which can be made accessible in open-science platforms for community based performance evaluation, further quantification, or the development of advanced methods for filtering and inversion.  The fact that this approach can be operated in an imaging mode, at standard scan rates, effectively opens the door for meaningful material investigations on the influence of SPV on structural components under different excitation conditions (e.g., wavelength and light intensity dependence). We have demonstrated the usefulness of adoption of data science clustering tools for high dimensional datasets afforded by this method, indeed the emerging trend of such techniques in the SPM community brings forward the challenge of adoption of such tools. Methods ranging from multivariate statistical approaches (e.g.  principle component analysis), unsupervised clustering algorithms, to more advanced machine learning methods will become critical to provide insight particularly for the large volumes of SPM data becoming available. The analysis shown here can be further extended to included investigation of decay times in appropriate samples, which in turn can provide information on carrier diffusion length scales. In the future this temporally resolved approach may prove useful for detection and imaging of electron spin relaxation in quantum computing or spintronic devices, domain wall motion in ferromagnetics. Finally, it is likely that combining this approach with bottom-up illumination will be advantageous for exploration of real solar cell devices while avoiding potential complications arising from tip-induced shadow effects, and enabling multimodal imaging with existing optical microscopies (e.g. photoluminence).

**MATERIALS AND METHODS**

Measurements were performed on a commercial AFM microscope (Cypher ES, Asylum research and Oxford Instr. Comp.) equipped with a laser diode (Blue drive, $\lambda$ = 405 nm) used to excite a photoresponse in the sample. The integrated laser was used to excite an area of the sample directly under the tip with a power output of ~10 µW. Although it was not possible to precisely calibrate the illuminated area under the tip, we estimate the light power density to be ~ 10000 mW/cm$^2$. All measurements were performed using conductive Ti/Ir-coated AFM probes (ASYELEC, Electrolever from asylum research) with nominal mechanical resonance frequency and spring constant of 70 kHz and 2.0 N/m, respectively. Importantly these probes were chosen due there is design feature including visible tip geometry necessary for top-down illumination of the tip-sample region. For G-Mode imaging we used Matlab and LabView software for control of the data acquisition platform and for post processing. National Instruments PXIe-1073 coupled with the NI PXIe-6214 DAQ architecture was used to generate the excitation waveforms to both the excitation laser and tip voltage, as well as to capture the photodetector signal. The MAPbBr$_3$ studied in this work was fabricated with a structure of ITO/PEDOT: PSS/ MAPbBr$_3$. The poly(3,4-ethylenedioxythiophene) polystyrene sulfonate (PEDOT:PSS) which were spin coated on the cleaned indium tin oxide (ITO) glass substrates at 4000 rpm for 60 s and then annealed at 150 °C for 30 min on a hot plate. The MAPbBr$_3$ films was prepared by dissolving 0.8 mmol PbBr$_2$, 1.4 mmol MABr and 0.2 mmol lead acetate trihydrate (Pb(Ac)$_2$·3H$_2$O) in 1 mL Dimethylformamide (DMF) solution and stirred overnight. The mixed solution was spin-coated on ITO/PEDOT:PSS substrate in nitrogen atmosphere at the rate of 3000 rpm to form MAPbBr$_3$ film, then annealed for 30 minutes at 60 °C.

**AUTHOR INFORMATION**

Corresponding Author

Correspondence should be addressed to Liam Collins (email: collinslf@ornl.gov) and Sergei Kalinin (email: sergei2@ornl.gov)

Author Contributions

L.C., S.J., S.V.K., designed the study. S.J. developed the G-Mode acquisition while L.C applied it to KPFM. L.C. performed the experiments and wrote the analysis code. M.A, and J. Q. fabricated the samples. L.C, S.V.K. M.A. analysed the data and interpreted the results. L.C., M.A. and S.V.K. wrote the paper. All authors commented on the manuscript.

**FUNDING SOURCES**

The authors declare no competing financial interest

**ACKNOWLEDGEMENTS**

Research was conducted at and supported by the Center for Nanophase Materials Sciences, which is a DOE Office of Science User Facility. M.A, J.Q and B.H acknowledge the support from Air Force Office of Scientific Research (AFOSR) (FA 9550-15-1-0064), AOARD (FA2386-15-1-4104), and National Science Foundation (CBET-1438181).

# References

1. Green, M. A.; Emery, K.; Hishikawa, Y.; Warta, W.; Dunlop, E. D., Solar Cell Efficiency Tables (Version 45). *Progress in photovoltaics: research and applications* **2015,** *23*, 1-9.
2. Veldhuis, S. A.; Boix, P. P.; Yantara, N.; Li, M.; Sum, T. C.; Mathews, N.; Mhaisalkar, S. G., Perovskite Materials for Light‐Emitting Diodes and Lasers. *Advanced Materials* **2016,** *28*, 6804-6834.
3. Green, M., Perovskite Single-Junction and Silicon-or Cigs-Based Tandem Solar Cells: Hype or Hope? *Proc. of 29th EU PVSEC 2014* **2014**.
4. Stranks, S. D.; Snaith, H. J., Metal-Halide Perovskites for Photovoltaic and Light-Emitting Devices. *Nature nanotechnology* **2015,** *10*, 391.
5. Tan, Z.-K.; Moghaddam, R. S.; Lai, M. L.; Docampo, P.; Higler, R.; Deschler, F.; Price, M.; Sadhanala, A.; Pazos, L. M.; Credgington, D., Bright Light-Emitting Diodes Based on Organometal Halide Perovskite. *Nature nanotechnology* **2014,** *9*, 687-692.
6. Brenner, T. M.; Egger, D. A.; Kronik, L.; Hodes, G.; Cahen, D., Hybrid Organic—Inorganic Perovskites: Low-Cost Semiconductors with Intriguing Charge-Transport Properties. *Nature Reviews Materials* **2016,** *1*, 15007.
7. Tirmzi, A. M.; Dwyer, R. P.; Hanrath, T.; Marohn, J. A., Coupled Slow and Fast Charge Dynamics in Cesium Lead Bromide Perovskite. *ACS Energy Letters* **2017,** *2*, 488-496.
8. Sanchez, R. S.; Gonzalez-Pedro, V.; Lee, J.-W.; Park, N.-G.; Kang, Y. S.; Mora-Sero, I.; Bisquert, J., Slow Dynamic Processes in Lead Halide Perovskite Solar Cells. Characteristic Times and Hysteresis. *The journal of physical chemistry letters* **2014,** *5*, 2357-2363.
9. Bergmann, V. W.; Weber, S. A.; Ramos, F. J.; Nazeeruddin, M. K.; Grätzel, M.; Li, D.; Domanski, A. L.; Lieberwirth, I.; Ahmad, S.; Berger, R., Real-Space Observation of Unbalanced Charge Distribution inside a Perovskite-Sensitized Solar Cell. *Nature communications* **2014,** *5*, 5001.
10. Snaith, H. J.; Abate, A.; Ball, J. M.; Eperon, G. E.; Leijtens, T.; Noel, N. K.; Stranks, S. D.; Wang, J. T.-W.; Wojciechowski, K.; Zhang, W., Anomalous Hysteresis in Perovskite Solar Cells. *The journal of physical chemistry letters* **2014,** *5*, 1511-1515.
11. Xiao, Z.; Yuan, Y.; Shao, Y.; Wang, Q.; Dong, Q.; Bi, C.; Sharma, P.; Gruverman, A.; Huang, J., Giant Switchable Photovoltaic Effect in Organometal Trihalide Perovskite Devices. *Nature materials* **2015,** *14*, 193-198.
12. Yuan, Y.; Li, T.; Wang, Q.; Xing, J.; Gruverman, A.; Huang, J., Anomalous Photovoltaic Effect in Organic-Inorganic Hybrid Perovskite Solar Cells. *Science Advances* **2017,** *3*, e1602164.
13. Vorpahl, S. M.; Stranks, S. D.; Nagaoka, H.; Eperon, G. E.; Ziffer, M. E.; Snaith, H. J.; Ginger, D. S., Impact of Microstructure on Local Carrier Lifetime in Perovskite Solar Cells. *Science* **2015**, aaa5333.
14. Guo, Z.; Manser, J. S.; Wan, Y.; Kamat, P. V.; Huang, L., Spatial and Temporal Imaging of Long-Range Charge Transport in Perovskite Thin Films by Ultrafast Microscopy. *Nature communications* **2015,** *6*, 7471.
15. Shao, Y.; Fang, Y.; Li, T.; Wang, Q.; Dong, Q.; Deng, Y.; Yuan, Y.; Wei, H.; Wang, M.; Gruverman, A., Grain Boundary Dominated Ion Migration in Polycrystalline Organic–Inorganic Halide Perovskite Films. *Energy & Environmental Science* **2016,** *9*, 1752-1759.
16. Yang, B.; Brown, C. C.; Huang, J.; Collins, L.; Sang, X.; Unocic, R. R.; Jesse, S.; Kalinin, S. V.; Belianinov, A.; Jakowski, J., Enhancing Ion Migration in Grain Boundaries of Hybrid Organic–Inorganic Perovskites by Chlorine. *Advanced Functional Materials* **2017**.

17. Nonnenmacher, M.; o'Boyle, M.; Wickramasinghe, H. K., Kelvin Probe Force Microscopy. *Applied physics letters* **1991,** *58*, 2921-2923.
18. Melitz, W.; Shen, J.; Kummel, A. C.; Lee, S., Kelvin Probe Force Microscopy and Its Application. *Surface Science Reports* **2011,** *66*, 1-27.
19. Berger, R.; Butt, H. J.; Retschke, M. B.; Weber, S. A., Electrical Modes in Scanning Probe Microscopy. *Macromolecular rapid communications* **2009,** *30*, 1167-1178.
20. Shikler, R.; Fried, N.; Meoded, T.; Rosenwaks, Y., Measuring Minority-Carrier Diffusion Length Using a Kelvin Probe Force Microscope. *Physical Review B* **2000,** *61*, 11041.
21. Loppacher, C.; Zerweck, U.; Teich, S.; Beyreuther, E.; Otto, T.; Grafström, S.; Eng, L. M., Fm Demodulated Kelvin Probe Force Microscopy for Surface Photovoltage Tracking. *Nanotechnology* **2005,** *16*, S1.
22. Palermo, V.; Palma, M.; Samorì, P., Electronic Characterization of Organic Thin Films by Kelvin Probe Force Microscopy. *Advanced materials* **2006,** *18*, 145-164.
23. Henning, A.; Günzburger, G.; Jöhr, R.; Rosenwaks, Y.; Bozic-Weber, B.; Housecroft, C. E.; Constable, E. C.; Meyer, E.; Glatzel, T., Kelvin Probe Force Microscopy of Nanocrystalline Tio2 Photoelectrodes. *Beilstein journal of nanotechnology* **2013,** *4*, 418.
24. Tennyson, E. M.; Garrett, J. L.; Frantz, J. A.; Myers, J. D.; Bekele, R. Y.; Sanghera, J. S.; Munday, J. N.; Leite, M. S., Nanoimaging of Open‐Circuit Voltage in Photovoltaic Devices. *Advanced Energy Materials* **2015,** *5*.
25. Shao, G.; Glaz, M. S.; Ma, F.; Ju, H.; Ginger, D. S., Intensity-Modulated Scanning Kelvin Probe Microscopy for Probing Recombination in Organic Photovoltaics. *ACS nano* **2014,** *8*, 10799-10807.
26. Kronik, L.; Shapira, Y., Surface Photovoltage Phenomena: Theory, Experiment, and Applications. *Surface Science Reports* **1999,** *37*, 1-206.
27. Schumacher, Z.; Miyahara, Y.; Spielhofer, A.; Grutter, P., Measurement of Surface Photovoltage by Atomic Force Microscopy under Pulsed Illumination. *Physical Review Applied* **2016,** *5*, 044018.
28. Yuan, Y.; Huang, J., Ion Migration in Organometal Trihalide Perovskite and Its Impact on Photovoltaic Efficiency and Stability. *Accounts of chemical research* **2016,** *49*, 286-293.
29. Cojocaru, L.; Uchida, S.; Tamaki, K.; Jayaweera, P. V.; Kaneko, S.; Nakazaki, J.; Kubo, T.; Segawa, H., Determination of Unique Power Conversion Efficiency of Solar Cell Showing Hysteresis in the Iv Curve under Various Light Intensities. *Scientific reports* **2017,** *7*, 11790.
30. Coffey, D. C.; Ginger, D. S., Time-Resolved Electrostatic Force Microscopy of Polymer Solar Cells. *Nature materials* **2006,** *5*, 735-740.
31. Giridharagopal, R.; Rayermann, G. E.; Shao, G.; Moore, D. T.; Reid, O. G.; Tillack, A. F.; Masiello, D. J.; Ginger, D. S., Submicrosecond Time Resolution Atomic Force Microscopy for Probing Nanoscale Dynamics. *Nano letters* **2012,** *12*, 893-898.
32. Karatay, D. U.; Harrison, J. S.; Glaz, M. S.; Giridharagopal, R.; Ginger, D. S., Fast Time-Resolved Electrostatic Force Microscopy: Achieving Sub-Cycle Time Resolution. *Review of Scientific Instruments* **2016,** *87*, 053702.
33. Dwyer, R. P.; Nathan, S. R.; Marohn, J. A., Microsecond Photocapacitance Transients Observed Using a Charged Microcantilever as a Gated Mechanical Integrator. *Science Advances* **2017,** *3*, e1602951.
34. Luria, J. L.; Schwarz, K. A.; Jaquith, M. J.; Hennig, R. G.; Marohn, J. A., Spectroscopic Characterization of Charged Defects in Polycrystalline Pentacene by Time‐ and Wavelength‐Resolved Electric Force Microscopy. *Advanced Materials* **2011,** *23*, 624-628.

35. Garrett, J. L.; Tennyson, E. M.; Hu, M.; Huang, J.; Munday, J. N.; Leite, M. S., Real-Time Nanoscale Open-Circuit Voltage Dynamics of Perovskite Solar Cells. *Nano Letters* **2017,** *17*, 2554-2560.
36. Belianinov, A.; Kalinin, S. V.; Jesse, S., Complete Information Acquisition in Dynamic Force Microscopy. *Nature communications* **2015,** *6*, 6550.
37. Collins, L.; Belianinov, A.; Somnath, S.; Rodriguez, B. J.; Balke, N.; Kalinin, S. V.; Jesse, S., Multifrequency Spectrum Analysis Using Fully Digital G Mode-Kelvin Probe Force Microscopy. *Nanotechnology* **2016,** *27*, 105706.
38. Collins, L.; Belianinov, A.; Somnath, S.; Balke, N.; Kalinin, S. V.; Jesse, S., Full Data Acquisition in Kelvin Probe Force Microscopy: Mapping Dynamic Electric Phenomena in Real Space. *Scientific reports* **2016,** *6*, 30557.
39. Collins, L.; Ahmadi, M.; Wu, T.; Hu, B.; Kalinin, S. V.; Jesse, S., Breaking the Time Barrier in Kelvin Probe Force Microscopy: Fast Free Force Reconstruction Using the G-Mode Platform. *ACS nano* **2017,** *11*, 8717-8729.
40. Collins, L.; Belianinov, A.; Proksch, R.; Zuo, T.; Zhang, Y.; Liaw, P. K.; Kalinin, S. V.; Jesse, S., G-Mode Magnetic Force Microscopy: Separating Magnetic and Electrostatic Interactions Using Big Data Analytics. *Applied Physics Letters* **2016,** *108*, 193103.
41. Nie, W.; Blancon, J.-C.; Neukirch, A. J.; Appavoo, K.; Tsai, H.; Chhowalla, M.; Alam, M. A.; Sfeir, M. Y.; Katan, C.; Even, J., Light-Activated Photocurrent Degradation and Self-Healing in Perovskite Solar Cells. *Nature communications* **2016,** *7*, 11574.
42. Moerman, D.; Eperon, G. E.; Precht, J. T.; Ginger, D. S., Correlating Photoluminescence Heterogeneity with Local Electronic Properties in Methylammonium Lead Tribromide Perovskite Thin Films. *Chemistry of Materials* **2017,** *29*, 5484-5492.
43. Zhang, T.; Cheung, S. H.; Meng, X.; Zhu, L.; Bai, Y.; Ho, C. H. Y.; Xiao, S.; Xue, Q.; So, S. K.; Yang, S., Pinning Down the Anomalous Light Soaking Effect toward High-Performance and Fast-Response Perovskite Solar Cells: The Ion-Migration-Induced Charge Accumulation. *The journal of physical chemistry letters* **2017,** *8*, 5069-5076.
44. Zhao, C.; Chen, B.; Qiao, X.; Luan, L.; Lu, K.; Hu, B., Revealing Underlying Processes Involved in Light Soaking Effects and Hysteresis Phenomena in Perovskite Solar Cells. *Advanced Energy Materials* **2015,** *5*.
45. Gottesman, R.; Zaban, A., Perovskites for Photovoltaics in the Spotlight: Photoinduced Physical Changes and Their Implications. *Accounts of chemical research* **2016,** *49*, 320-329.
46. Maturova, K.; Kemerink, M.; Wienk, M. M.; Charrier, D. S.; Janssen, R. A., Scanning Kelvin Probe Microscopy on Bulk Heterojunction Polymer Blends. *Advanced Functional Materials* **2009,** *19*, 1379-1386.
47. Kodinariya, T. M.; Makwana, P. R., Review on Determining Number of Cluster in K-Means Clustering. *International Journal* **2013,** *1*, 90-95.